\documentclass[12pt,onecolumn]{IEEEtran}
\usepackage{setspace}
\usepackage[utf8]{inputenc} % allow utf-8 input
\usepackage[T1]{fontenc}    % use 8-bit T1 fonts
\usepackage{hyperref}       % hyperlinks
\usepackage{url}            % simple URL typesetting
\usepackage{booktabs}       % professional-quality tables
\usepackage{amsfonts}       % blackboard math symbols
\usepackage{nicefrac}       % compact symbols for 1/2, etc.
\usepackage{microtype}      % microtypography
\usepackage{xcolor}         % colors
\usepackage{color}
\usepackage{subcaption}
\usepackage{caption}

\usepackage{silence}
\usepackage{cite}
\usepackage{amsmath}      % For math symbols and align environments
\usepackage{amsfonts}     % For math fonts like \mathbb
\usepackage{algorithm}    % For creating algorithms
\usepackage{algpseudocode} % For algorithmic environment

\usepackage{amssymb}
\usepackage{amsmath}
\usepackage{multirow}
\usepackage{multicol}   %Columns
\usepackage[justification=centering]{caption}
\usepackage{graphicx}

\usepackage{caption}
\usepackage{subcaption}

\usepackage{ifthen}
\newboolean{showcomments}
\setboolean{showcomments}{true}

\newcommand{\note}[1]{\ifthenelse{\boolean{showcomments}} {\textcolor{red}{#1}}{}}

\usepackage{epstopdf,float,amsfonts}
\usepackage{epsfig}

\newtheorem{myth}{Theorem}
\newtheorem{mylemma}{Lemma}
\newtheorem{mycorollary}{Corollary}
\newtheorem{myassumption}{Assumption}
\newcommand{\qed}{\hfill \ensuremath{\square}}

\title{On the System Theoretic Offline Learning of Continuous-Time LQR with Exogenous Disturbances}
\usepackage{times}
\author{\IEEEauthorblockN{ Sayak Mukherjee, Ramij R. Hossain, Mahantesh Halappanavar
}
\thanks{Authors are with the Pacific Northwest National Laboratory, Richland, WA 99352, USA, Email: sayak.mukherjee@pnnl.gov. }
}
\begin{document}

\maketitle

%\usepackage[preprint]{neurips_2023}
% to compile a preprint version, e.g., for submission to arXiv, add add the
% [preprint] option:
%     \usepackage[preprint]{neurips_2023}

% to compile a camera-ready version, add the [final] option, e.g.:
%     \usepackage[final]{neurips_2023}

% to avoid loading the natbib package, add option nonatbib:
%    \usepackage[nonatbib]{neurips_2023}

% The \author macro works with any number of authors. There are two commands
% used to separate the names and addresses of multiple authors: \And and \AND.
%
% Using \And between authors leaves it to LaTeX to determine where to break the
% lines. Using \AND forces a line break at that point. So, if LaTeX puts 3 of 4
% authors names on the first line, and the last on the second line, try using
% \AND instead of \And before the third author name.

  % examples of more authors
  % \And
  % Coauthor \\
  % Affiliation \\
  % Address \\
  % \texttt{email} \\
  % \AND
  % Coauthor \\
  % Affiliation \\
  % Address \\
  % \texttt{email} \\
  % \And
  % Coauthor \\
  % Affiliation \\
  % Address \\
  % \texttt{email} \\
  % \And
  % Coauthor \\
  % Affiliation \\
  % Address \\
  % \texttt{email} \\

\begin{abstract}
  We analyze offline designs of linear quadratic regulator (LQR) strategies with uncertain disturbances. First, we consider the scenario where the exogenous variable can be estimated in a controlled environment, and subsequently, consider a more practical and challenging scenario where it is unknown in a stochastic setting. Our approach builds on the fundamental learning-based framework of adaptive dynamic programming (ADP), combined with a Lyapunov-based analytical methodology to design the algorithms and derive sample-based approximations motivated from the Markov decision process (MDP)-based approaches. For the scenario involving non-measurable disturbances, we further establish stability and convergence guarantees for the learned control gains under sample-based approximations. The overall methodology emphasizes simplicity while providing rigorous guarantees. Finally, numerical experiments focus on the intricacies and validations for the design of offline continuous-time LQR with exogenous disturbances.
\end{abstract}

\section{Introduction}

% \subsection{Related Work}

In recent years we have seen advancement of reinforcement learning \cite{ RL, silver2016mastering, schulman2017proximal} in both discrete and continuous control tasks. Different forms of formulations using Markov decision processes \cite{RL} and also using system theoretic state space based approaches \cite{vrabie1, jiang1, mukherjee2021reduced, mukherjee2022reinforcement} are introduced. In many of the systems, we often encounter uncertainty and stochasticity which makes the continuous control tasks more complex. In the MDP-based frameworks, the stochasticity in the dynamics is considered to be an inherent assumption, however, the system theoretic approaches require special mathematical treatments to provide stability and convergence guarantees when dealing with uncertainties. In the control theoretic approach, both linear and non-linear optimal control problems are solved with a data-driven approaches to solve algebraic Riccati equations (AREs) or Hamilton-Jacobi-Bellman (HJB) equations \cite{jiang_assumption}. Moreover, works such as \cite{depersis2020formulas} have bring forth ideas from Willems' fundamental lemma \cite{willems2005note} and behavioral system theory. Variations of such works have investigated robust designs in presence of uncertainties \cite{bisoffi2022data}. On the uncertain system works such as \cite{jiang2011approximate,bian2016adaptive, bian2018stochastic, cao2024analyzing} have considered the stochastic differential equation (SDE) based approach. However, due to coupled nature of the noise variable in the dynamics, many works have to assume some availability of measurements of such exogenous variables or the use of a controlled environment. Model-free random search based algorithms to solve linear quadratic regulator (LQR) problems are also proposed in \cite{fazel2018global, mohammadi2020learning, mania2018simple} although these approaches utilize an online controller-in-the-loop methodology. 

On the parallel, recent efforts have been on developing offline RL approaches, where the controller do not require to interact with the environment online, and utilize previously collected batch interaction data. In the MDP framework, one set of algorithms work with constraining the policies to be close to the behavioral policies (e.g., BCQ (Batch-Constrained Q-learning) \cite{fujimoto2019bcq}), or with value function regularization such as penalizing the Q-function when assigning high values to unseen out-of-distribution actions, resulting into conservative solution (Conservative Q-learning (CQL) \cite{kumar2020cql}). \cite{levine2020offline} reviews the progress and challenges associated with offline RL from the MDP viewpoint. To extend the idea of offline RL solution for the system theoretic state space setting, we merge few ideas from integral reinforcement learning and sample based solutions used in MDP setting to tackle uncertainty.  

\noindent \textit{Contributions.} We analyze both of the offline LQR strategies when the exogenous variable is estimated in a controlled environment, and also when it is unknown in stochastic environment. We utilize the fundamental learning-based strategy of adaptive dynamic programming (ADP) along with Lyapunov-based analytical framework to develop the algorithms and formulating approximations using sample-based solutions as frequently considered in MDP-based framework. Thereafter, for the scenario with non-measurable unknown disturbance, we perform the stability and convergence analysis of the learned control gain solution using the sample-based approximations. The methodology focuses on providing simplicity along with mathematical guarantees. Our numerical experiments also validates the algorithmic approaches and assumptions.  

The rest of the paper is organized as follows. Section II describes the problem formulations and background on model-based planning and model-free approaches. Section III then delves deep into the methodology and analysis with main focus on the learning in uncertain environment, and developing the approximated episodic offline policy learning. Numerical example is described in Section IV with concluding remarks given in Section V.

\section{Model and Problem Formulation}
We consider a linear time-invariant (LTI) continuous-time dynamical system perturbed with exogenous inputs:
\begin{align}\label{system}
    \dot{x} = Ax + Bu + Ee,\; x(0)=x_0,
\end{align}
where $x \in \mathbb{R}^{n}, u \in \mathbb{R}^{m}$ are the states and control inputs, and $e \in \mathbb{R}^{p} $ is the exogenous inputs to the system which may cause stochasticity in the dynamics. The linear quadratic regulator problem considers infinite horizon cost minimization of the form:
\begin{align}
    \label{Ji}
   \text{min}_{u} J (x(0),u) = \mathbb{E}_\mathcal{T} \{ [\int_{0}^{\infty} (x(\tau)^TQx(\tau) + u(\tau)^TRu(\tau)) d\tau],
\end{align}
and the control policy is given by the state-feedback law as $u= -Kx$, where $K \in \mathbb{R}^{m\times n}$. Here $Q$ and $R$ are positive semi-definite and definite matrices that puts penalty in state fluctuations and control excursions to improve both state and control performances for the dynamics in closed-loop. For the existance of optimal solution we make the following assumption as considered standard in the literature:
\begin{myassumption}
    The pair $(A,B)$ is stabilizable and $(Q^{1/2}, A)$  is detectable \cite{zhou1998essentials}.  
\end{myassumption} 
\par

% \subsection{LQR Methods Review}
\noindent \textit{Model-based planning:} Conventionally when the dynamics is known for the nominal system $\dot{x} = Ax + Bu$ we can perform model-based planning. In the context of the infinite horizon LQR problem without noise, planning can be accomplished through the solution of the Algebraic Riccati Equation (ARE):
\begin{align}
&A^TP + PA + Q -PBR^{-1}B^TP = 0, \label{pol_eval}\\
&K = R^{-1}B^TP. \label{pol_update}
\end{align}
Here $P$ is a postive definite matrix that parametrizes the cost-to-go from the action taken. There are vast literature in solving Riccati equations stemming from iterative solvers such as Kleinman's algorithm \cite{kleinman} to Hamiltonian approaches and Semi-definite programs \cite{boyd1994lmi}. \par

% \textbf{Zeroth-order gradients - model-free random search:} 
% Recently works such as \cite{} framed this problem to a classical gradient-based optimization viewpoint for discrete-time and continuous-time cases. These works have considered the dynamics $\dot{x} = Ax + Bu$ with stochastic initial conitions, i.e., $x_0 \sim \mathcal{D}$ for some distribution $\mathcal{D}$. Thereafter, policy gradient methods implement gradient-descent for the policy as:
% \begin{align}
%     K \leftarrow K - \eta \nabla J(K),
% \end{align}
% where the step-size is denoted as $\eta$. The gradient of the cost $\nabla J(K)$ for the model-free approaches are generally computed by performing random search on a dynamic oracle by perturbing the initial policies by $K + rU$, where $U$ is sampled from random vector uniformly distributed on the sphere, and $r$ is a smoothing constant. This can be performed for one point gradient estimate or two-point gradient estimate with $K \pm rU$ as in \cite{}. For the two-point estimate the zeroth-order approximated gradient is computed after running simulations using the dynamic oracle:
% \begin{align}
%     \hat{\nabla J(K)} = \frac{1}{2rN} \sum_{i=1}^{N} (\hat{J}_i(K + rU) - \hat{J}_i(K - rU)) ,
% \end{align}
% where the state and control trajectories are stored for $N$ episodes of considerate lengths.  

For this work we consider the model-free setting and make the following assumption.
\begin{myassumption}
    The dynamic state matrix $A$ is unknown, although the values of $n$, and $m$ are known.
\end{myassumption}
\par
With this unknown state dynamics, we would like to learn an optimal feedback policy $u=-Kx$ for the system \eqref{system} satisfying assumptions $1, 2$, such that the closed-loop system is stable and the objective \eqref{Ji} is minimized. Recently works such as \cite{fazel2018global, mohammadi2020learning} framed this problem to a classical gradient-based optimization viewpoint for discrete-time and continuous-time cases. These works have considered the dynamics $\dot{x} = Ax + Bu$ with stochastic initial conitions, i.e., $x_0 \sim \mathcal{D}$ for some distribution $\mathcal{D}$. Thereafter, policy gradient methods implement gradient-descent for the policy as:
  $  K \leftarrow K - \eta \nabla J(K),
$
where the step-size is denoted as $\eta$. The gradient of the cost $\nabla J(K)$ for the model-free approaches are generally computed by performing random search on a dynamic oracle by perturbing the initial policies by $K + rU$, where $U$ is sampled from random vector uniformly distributed on the sphere, and $r$ is a smoothing constant. However, this results into an \textit{online control update algorithm}. We, on the other hand, are interested in offline LQR solutions where we gather different set of trajectories using the oracle model and subsequently compute a feasible and optimal control gain.  Therefore, we will consider a different route than performing random search based optimization, motivated from the classical Riccati equation solution based approaches, adaptive dynamic programming, and Lyapunov considerations. Works such as \cite{jiang1, vrabie1} contributed to these approaches for the deterministic LQR problems. In the context of uncertain systems, several works (e.g., \cite{jiang2011approximate,bian2016adaptive,bian2018stochastic, cao2024analyzing}) have adopted a stochastic differential equation (SDE) framework. However, because the noise terms appear coupled with the system dynamics in the learning algorithms, these approaches often rely on the assumption that certain measurements of the exogenous disturbances are available.\\
\noindent \textit{Problem Statement: Utilize offline trajectories of \eqref{system} to learn an optimal feedback policy $u=-Kx$ satisfying assumptions $1, 2$, such that the closed-loop system is stable and the objective \eqref{Ji} is minimized. }

\section{Methodology and Analysis}

We first start by considering a controlled environment with measurable exogenous inputs for better exposition, with the main contribution is presented in the subsequent sub-section where the learning is performed in an uncertain environment. 
\subsection{Learning in Controlled Environment: Exact One-shot Offline Learning}

We first consider the scenario where the learning experiment has been conducted in a controlled environment, thus making the exogeneous inputs $e(t)$ measurable. Apart from that there may be physical processes where the exogenous inputs can be estimated.
% If the controller were to use $y(t)$ for feedback then it would find $u = -Ky(t)$ satisfying
% \begin{align}\label{J_red}
% \text{minimize\;\;}&\bar{J}(y(0);u)= \int_0^{\infty} (y^T Q y + u^T R u )dt, \\
% & \mbox{s.t.} \;\; A - BK \in \mathbb{RH}_{\infty}.
% \end{align}
The model-based optimal solution for the control $u(t) = -Kx(t)$ pertaining to \eqref{system} without any exogenous inputs, i.e., $e(t) = 0$ is given in \eqref{pol_eval} with $P \succ 0$. 
% \begin{align}
% &A^TP + PA + Q -PBR^{-1}B^TP = 0, \label{pol_eval}\\
% &K = R^{-1}B^TP. \label{pol_update}
% \end{align}
Here $P$ is the cost-matrix that performs \textit{policy evaluation} using \eqref{pol_eval}, and \eqref{pol_update} performs policy update step. This is also analogous to \textit{actor-critic} updates where $P$ is the critic and the actor is the computation of $K$, and subsequently implemented as $u = -Kx$. The policy $K$ without noise can be iteratively computed using the model-based Kleinman's algorithm \cite{kleinman}, which we recall for this problem setup for better comprehension.\par
\begin{myth}
\label{Kleinman_power}
\textbf{\cite{kleinman}} \textit{Let $K_0  $ be such that $A-B  K_0$ is Hurwitz. Then, for $k=0,1,\dots$ \\
1. Solve for $P_k$ starting with stabilizing $K_0$ (Policy Evaluation):
\begin{align}\label{Kleinman1 ps}
\hspace{-.5 cm} A_{k}^TP_k + P_kA_{k} + Q + K_k^TRK_k = 0, A_{k} = A-BK_k.
\end{align}
2. Update feedback matrix (Policy update) :
\begin{align}\label{Kleinman2 ps}
K_{k+1} = R^{-1}B^TP_k.
\end{align}
Then $A-B  K$ is Hurwitz and $K_k$ and $P_k$ converges to optimal $K,P$ as $k \rightarrow \infty$.}  \qed
\end{myth}
Based on this theorem, one can develop an iterative exact policy learning algorithm by compensating the impact of exogenous variable $e(t)$. The oracle has been explored by injecting exploration signal $u_0$ for sufficiently large time-steps. We would like to have this exploration in such a manner that the dynamics can be persistently excited. Persistency of excitation is a staple condition in the literature of adaptive control, and can be ensured by sufficient number of samples such that the rank of the matrix composed of $\mathcal{I}_{xx}$, $\mathcal{I}_{xu_0}$, and $\mathcal{I}_{xe}$ equals $\frac{n(n+1)}{2} + nm + np. $ We denote two quantities using vectors $p(t), q(t)$ as follows ($\otimes$ denote Kronecker product).
\begin{align}
    &\mathcal{D}_{pp} = \begin{bmatrix}
p \otimes p |_{t_1}^{t_1+T} ,& \dots & p \otimes p |_{t_l}^{t_l+T}
\end{bmatrix}^T,\\
&\mathcal{I}_{pq} = \begin{bmatrix}
\int_{t_1}^{t_1+T}(p \otimes q) d\tau ,& \dots & \int_{t_l}^{t_l+T} (p \otimes q) d\tau \\
\end{bmatrix} ^T.   
\end{align}
\begin{mylemma}
 With persistently excited oracle, the states and the policies are coupled by the following trajectory based relationship: 
\begin{align}\label{eq:update_exact}
 \underbrace{\begin{bmatrix}
\mathcal{D}_{xx}  -2\mathcal{I}_{xx}(I \otimes K_k^TR)  -2\mathcal{I}_{xu_0}(I \otimes R)  -2\mathcal{I}_{xe}
\end{bmatrix}}_{\Theta_k}
&\begin{bmatrix}
vec(P_k) \\ vec(K_{k+1}) \\ vec(B^TP_k)
\end{bmatrix}  =\underbrace{-\mathcal{I}_{xx}vec(Q_k)}_{\Phi_k}.
\end{align}
\normalsize
\end{mylemma}

Proof: By writing \eqref{system} incorporating $u = -K_kx$ as follows,
\begin{align}
\dot{x} &= Ax + B\tilde{u} + Ee,\\
&= (A - BK_k)x + B(K_kx + u) + Ee.
\end{align} 
Given $\tilde{u} = u_0$, using a sufficiently exciting control policy maintaining bounded system states during the learning's exploration phase \cite{jiang_book}, we have:
% The control policy $u_0$ can be arbitrary, as long as the system states remain bounded \cite{jiang_book}. For example, following \cite{jiang_book} one can design $u_0$ as a sum of sinusoids.
\begin{align}
\frac{d}{dt}(x^TP_kx) &= x^T(A_{k}^TP_k + P_kA_{k})x + 2(K_kx+u_0)^TB^TP_kx \nonumber \\ & \;\;\;\; + 2(e^TE^TP)x, \\ 
%  &= -x^TQ_kx + 2(K_kx+u_0)^TB^TP_kx + 2(v^TD^TP)x\\
&= -x^T\bar{Q}_kx + 2(K_kx+u_0)^TRK_{k+1}x + 2(e^TE^TP)x, \nonumber
\end{align}
\normalsize
where, $\bar{Q}_k = Q + K_k^TRK_k$. This results into the trajectory-defined coupled equation as:
\begin{align}
\label{main eqn ps}
 &\hspace{-.3 cm} x^T(t+T)P_kx(t+T) - x^T P_k x\nonumber  
 -  2\int_{t}^{t+T}((K_kx+u_0)^TR K_{k+1}x )d\tau  \nonumber \\ & \;\;\;\;\; = -\int_{t}^{t+T}(x^T \bar{Q}_k x + 2(e^TE^TP)x ) d\tau.
\end{align}
\normalsize
Using $x^T\bar{Q}_kx = (x^T \otimes x^T)\,\mbox{vec}(\bar{Q}_k), e^TE^TPx = (x^T \otimes e^T)\,\mbox{vec}(E^TP)$, the lemma is proven. \qed

The offline learning algorithm can be derived by formulating an iterative form of \eqref{eq:update_exact}, utilizing measurements of $x(t)$, $u_0(t)$, and $e(t)$ obtained from the exploration, as outlined in Algorithm 1.
\begin{algorithm}
\label{alg:controlled}

\caption{Exact One-shot Offline Learning}
\begin{algorithmic}[1]
\State \textbf{Offline data generation:} Store data \((x, e, u_0)\) over the time intervals \( (t_1, t_2, \dots, t_l), \) where \( t_i - t_{i-1} = T \). 
\State Construct the matrices \( \mathcal{D}_{xx}, \mathcal{I}_{xx}, \mathcal{I}_{xu_0}, \mathcal{I}_{xe} \), ensuring that the rank condition \( \text{rank}(\mathcal{I}_{xx} \;\; \mathcal{I}_{xu_0}\;\; \mathcal{I}_{xe}) = \frac{n(n+1)}{2} + nm + np \) is satisfied.
\vspace{0.2cm}

\State \textbf{Policy Update Iteration:} Start with an initial stabilizing gain \( K_0 \), and update the policy iteratively:

\For{$k=0,1,\dots$}
\State Compute \( K_{k+1} \) using the following equation:
\begin{align}\label{eq:update_controlled}
\underbrace{
\begin{bmatrix}
\mathcal{D}_{xx} & - 2 \mathcal{I}_{xx}(I \otimes K_k^T R) - 2 \mathcal{I}_{xu_0}(I \otimes R) & - 2 \mathcal{I}_{xe}
\end{bmatrix}
}_{\Theta_k}
\begin{bmatrix}
\text{vec}(P_k) \\ \text{vec}(K_{k+1}) \\ \text{vec}(E^T P_k)
\end{bmatrix}  \nonumber \\ 
= \underbrace{- \mathcal{I}_{xx} \text{vec}(Q_k)}_{\Phi_k}.
\end{align}
\State Terminate when \( |P_k - P_{k-1}| < \varsigma \), where \( \varsigma > 0 \) is a small threshold.
\EndFor
\vspace{0.2cm}

\State \textbf{Applying \( K \) to the System:} Apply the control law \( u = -K x \) and remove \( u_0 \).
\end{algorithmic}
\end{algorithm}
\normalsize

% \noindent \textit{Remark 2:} For each $k=0,1,2,\cdots$ it is assumed that there exists a sufficiently large integer $l_k > 0$ signifying large enough sampling intervals, such that rank($\Theta_k$) = $r(r+1)/2 + rm + rp$. It follows from \cite{jiang_book} that this rank condition is satisfied by utilizing data from at least twice as many sampling intervals as the number of unknowns.\par
% \noindent \textit{Remark 2:} If $A$ in \eqref{eq:statecompact1} is Hurwitz, then the controller update iteration in \eqref{eq:update} can be started without any stabilizing initial control. Otherwise, stabilizing $K_0$ is required, as commonly encountered in the RL literature \cite{jiang_book}. This is mainly due to its equivalence with Kleinman's algorithm in Theorem $1$.\par 

\begin{myth}
\label{reduced_stability} \textit{Performing Algorithm $1$ using $x(t), u_0(t)$, and $e(t)$ will lead to stable and optimal $P$ and $K$ for \eqref{system} with $e(t) =0$. } 
 \end{myth}
\textit{Proof:} The iterative solution of \eqref{eq:update_controlled} corresponds to Algorithm $1$, utilizing the exogenous variable $e(t)$. According to the formulation of \eqref{eq:update_controlled}, any solution $P_k$ from Theorem 1 satisfies the equation for the $k^{th}$ iteration as follows:

\begin{align}\label{step2 ps}
    \Theta_k \begin{bmatrix}
vec(P_k) \\ vec(K_{k+1}) \\ vec(E^TP_k)
\end{bmatrix} = \Phi_k.
\end{align}

If the rank of the matrix composed of $\mathcal{I}_{xx}$, $\mathcal{I}_{xu_0}$, and $\mathcal{I}_{xe}$ equals $\frac{n(n+1)}{2} + nm + np $, then $\Theta_k$ possesses full column rank. This implies that solutions for $P_k$ and $K_{k+1}$ will be unique as obtained from \eqref{step2 ps}. Consequently, these solutions correspond to those of \eqref{Kleinman1 ps}-\eqref{Kleinman2 ps}. Given this equivalence, it can be inferred that the offline RL in the controlled environment will converge to stable and optimal $P$ and $K$ of \eqref{system} when $e(t)=0$.\qed 

\subsection{Learning in Uncertain Environment: Approximated Episodic Policy Learning}

When the exogenous input is a noise signal, in most practical settings, the designer would not be able to measure it. Therefore, we assume the existence of an oracle that can generate trajectories, and formulate an approximated offline learning.

\begin{myassumption}
    The dynamic system's oracle is available to the designer, and therfore, we will have the following off-line genertaed data over $N$ different trajectories $\mathcal{T}$.
\begin{align}
    \mathcal{O} : \{ x_1^i, x_2^i, \dots, x_l^i, u_1^i, u_2^i, \dots, u_l^i \}_{i=1:N}
\end{align}
\end{myassumption}

Let us also consider a standard property for the noise signal.
\begin{myassumption}
    We consider a zero-mean Gaussian for the existing noise, i.e., $e(t) \sim \mathcal{N}(0, \Sigma)$. 
\end{myassumption}

For the episodic training setting, we can collect $N$ number of trajectories $\mathcal{T}$ with each trajectory sufficiently exciting in an offline way such that $\mathrm{rank}([\mathcal{I}_{xx} \;\; \mathcal{I}_{xu_0}]) = \tfrac{n(n+1)}{2} + nm$. The following Lemma gives an approximated expression of the trajectory relationship in a multi-episodic setup.

\noindent \begin{mylemma}
\label{approx_main_eq}
    The trajectory relationship for the states and inputs of the stochastic dynamic system \eqref{system} with the help of oracle $\mathcal{O}$ is given by:
    \begin{align}
        \label{main eqn ps}
 &\mathbb{E}_\mathcal{T} \{ x^T(t+T)P_kx(t+T) \} - \mathbb{E}_\mathcal{T}  \{x^T P_k x \} \nonumber \\  
 & -  2\mathbb{E}_\mathcal{T} \{ \int_{t}^{t+T}((K_kx+u_0)^TR K_{k+1}x )d\tau \}   = -\int_{t}^{t+T}\mathbb{E}_\mathcal{T}\{ (x^T \bar{Q}_k x ) d\tau\}.
    \end{align}
    \normalsize
\end{mylemma}

Proof: We can recall from the single-episode dynamical behavior we have,
\begin{align}
    \frac{d}{dt}(x^TP_kx) 
&= x^T(A_{k}^TP_k + P_kA_{k})x  \nonumber \\ & + 2(K_kx+u_0)^TRK_{k+1}x + 2(e^TE^TP)x, \nonumber
\end{align}
\normalsize

% let us consider a nominal dynamics for design:
% \begin{align}
% \dot{x} &= Ax + B\tilde{u} ,\\
% &= (A - BK_k)x + B(K_kx + u).
% \end{align} 

As the system is perturbed by noise we create multiple data realizations and perform expectation operation (sample-based estimates). 
%Considering $\tilde{u} = u_0$, an arbitrary control policy that keeps the system states bounded during the \textit{exploration} phase of the learning \cite{jiang_book}, we have
% The control policy $u_0$ can be arbitrary, as long as the system states remain bounded \cite{jiang_book}. For example, following \cite{jiang_book} one can design $u_0$ as a sum of sinusoids.
\begin{align}
 \mathbb{E}_\mathcal{T} \{\frac{d}{dt}(x^TP_kx) \} &= \mathbb{E}_\mathcal{T} \{x^T(A_{k}^TP_k + P_kA_{k})x \} +   2\mathbb{E}_\mathcal{T} \{e^TE^TPx \} \nonumber \\ 
 & \;\;\;\;\; +2\mathbb{E}_\mathcal{T} \{(K_kx+u_0)^TB^TP_kx \} \\ 
%  &= -x^TQ_kx + 2(K_kx+u_0)^TB^TP_kx + 2(v^TD^TP)x\\
 &= -\mathbb{E}_\mathcal{T} \{x^T\bar{Q}_kx \}+ 2\mathbb{E}_\mathcal{T} \{(K_kx+u_0)^TRK_{k+1}x \}, \nonumber
\end{align}
\normalsize
where, $\bar{Q}_k = Q + K_k^TRK_k$. Here we use the property that the process noise vector is independent of the state random variable at a particular time step, and we use the law of large numbers assuming we have sufficient number of realizations $N$ from the oracle $\mathcal{O}$, i.e., $\mathbb{E}_\mathcal{T} \{e^TE^TPx \} \to 0 $ at a time step with zero mean Gaussian process noise. Therefore, the trajectory based relationship will be given by,
\begin{align}
\label{main eqn ps uncertain}
 &\hspace{-.3 cm} \mathbb{E}_\mathcal{T} \{x^T(t+T)P_kx(t+T) \} - \mathbb{E}_\mathcal{T} \{x^T P_k x \} \nonumber \\  
 & -  2\mathbb{E}_\mathcal{T} \{\int_{t}^{t+T}((K_kx+u_0)^TR K_{k+1}x )d\tau \}  \nonumber \\ & \;\;\;\;\; = -\int_{t}^{t+T}\mathbb{E}_\mathcal{T} \{(x^T \bar{Q}_k x ) d\tau \}.
\end{align}
\normalsize
\qed

\begin{mycorollary}
Following Lemma \ref{approx_main_eq}, given the system dynamics with an exploration signal \( u_0(t) \) independent of \( x(t) \), the expected value of the change in quadratic costs over the interval \([t, t+T]\) is:

\begin{align}
\label{eq:covariance_full}
    &\text{Tr}(\Phi(T)^T P_k \Phi(T) \Sigma_x) + \mathbb{E}_\mathcal{T}(x)^T \Phi(T)^T P_k \Phi(T) \mathbb{E}_\mathcal{T}(x) 
    \nonumber \\ &- \left( \text{Tr}(P_k \Sigma_x) + \mathbb{E}_\mathcal{T}(x)^T P_k \mathbb{E}_\mathcal{T}(x) \right) \nonumber \\
    &\quad - 2 \int_t^{t+T} \left( \text{Tr}(K_k^T R K_{k+1} \Sigma_x) + \mathbb{E}_\mathcal{T}(x)^T K_k^T R K_{k+1} \mathbb{E}_\mathcal{T}(x) \right) d\tau \nonumber \\
    &= - \int_t^{t+T} \left( \text{Tr}(\bar{Q}_k \Sigma_x) + \mathbb{E}_\mathcal{T}(x)^T \bar{Q}_k E(x) \right) d\tau,
\end{align}
\normalsize
where, $\Sigma_x = \mathbb{E}[x(t) x(t)^T] $ (state covariance matrix) and $ 
    \Phi(T) = e^{A T}$  (state transition matrix).
\end{mycorollary}\qed

%Proof: 
% Using $x^T\bar{Q}_kx = (x^T \otimes x^T)\,\mbox{vec}(\bar{Q}_k)$,
% \begin{align}
%     y^T\bar{Q}_ky = (y^T \otimes y^T)vec(\bar{Q}_k)\nonumber, d^TE^TPy = (y^T \otimes d^T)vec(E^TP),
% \end{align}
%\subsubsection{Learning Algorithm}
To formulate the learning algorithm, one straightforward approach would be to consider the average state dynamics with sufficiently high number of episodes and applying law of large numbers leading to the dynamics $ \mathbb{E}_\mathcal{T} (\dot{x}) = A \mathbb{E}_\mathcal{T} (x) + B \mathbb{E}_\mathcal{T} (u) $, where $\mathbb{E}_\mathcal{T} (e) \to 0$. Therefore, from the data-driven Lyapunov-like treatment would lead to:
\begin{align}
\label{eq:average_state}
    &\left( \mathbb{E}_\mathcal{T} \left[ x(t+T) \right]^T P_k \mathbb{E}_\mathcal{T} \left[ x(t+T) \right] \right) 
    - \left( \mathbb{E}_\mathcal{T} \left[ x(t) \right]^T P_k \mathbb{E}_\mathcal{T} \left[ x(t) \right] \right) \nonumber \\
    &\quad - 2 \int_t^{t+T} \left( \left( K_k \mathbb{E}_\mathcal{T} \left[ x \right] + \mathbb{E}_\mathcal{T} \left[ u_0 \right] \right)^T R K_{k+1} \mathbb{E}_\mathcal{T} \left[ x \right] \right) d\tau 
    \nonumber \\
    & = - \int_t^{t+T} \left( \mathbb{E}_\mathcal{T} \left[ x \right]^T \bar{Q}_k \mathbb{E}_\mathcal{T} \left[ x \right] \right) d\tau
\end{align}
\normalsize
However, this would not be able to capture the state covariance, as using Corollary 1 and subtracting equation \eqref{eq:average_state} from \eqref{eq:covariance_full} would lead to:
\begin{align}
    \text{Tr}(P_k (\Sigma_x(t+T) - \Sigma_x(t))) 
    - 2 \int_t^{t+T} \text{Tr}(K_k^T R K_{k+1} \Sigma_x) d\tau \nonumber \\
    = - \int_t^{t+T} \text{Tr}(\bar{Q}_k \Sigma_x) d\tau,
\end{align}
\normalsize

 which will lead to error in gain computation. Therefore, we formulate the RL algorithm without the knowledge of $e(t)$ 
by formulating an iterative version of \eqref{main eqn ps}, using measurements of $x(t),u_0(t)$  over multiple realizations. The sample-based estimated averaged data matrices are given as following with algorithm 2 presents the approximated offline RL algorithm. 
\begin{align}
\label{sample_estimate}
    &\mathbb{E}_\mathcal{T} \{\mathcal{D}_{xx} \} \approx \begin{bmatrix}
\frac{1}{N} \sum_{i=1}^N (x^i \otimes x^i) |_{t_1}^{t_1+T} ,..,\frac{1}{N} \sum_{i=1}^N (x^i \otimes x^i) |_{t_l}^{t_l+T}
\end{bmatrix}^T,\\
\label{sample_estimate2}
& \mathbb{E}_\mathcal{T} \{\mathcal{I}_{xx} \} \approx \begin{bmatrix}
\frac{1}{N} \sum_{i=1}^N \int_{t_1}^{t_1+T}(x^i \otimes x^i) d\tau ,.., \frac{1}{N} \sum_{i=1}^N \int_{t_l}^{t_l+T} (x^i \otimes x^i) d\tau \\
\end{bmatrix} ^T.   \\
\label{sample_estimate3}
&\mathbb{E}_\mathcal{T} \{ \mathcal{I}_{xu_0}\} \approx \begin{bmatrix}
\frac{1}{N} \sum_{i=1}^N \int_{t_1}^{t_1+T}(x^i \otimes u_0^i) d\tau ,..,  \frac{1}{N} \sum_{i=1}^N \int_{t_l}^{t_l+T} (x^i \otimes u_0^i) d\tau \\
\end{bmatrix} ^T.   
\end{align}
\normalsize

\begin{algorithm}[ht]

\caption{Approximated Offline Learning}
\begin{algorithmic}[1]
\For{$i = 1 : N$}
    \State \textit{Offline Data:} Collect data ($x$ and $u_0$) for interval $(t_1, t_2, \ldots, t_m)$ with $t_i - t_{i-1} = T$.
    \State \textit{Construct} the matrices 
    $\mathcal{D}_{xx}, \mathcal{I}_{xx}, \mathcal{I}_{xu_0}$ 
    such that $\mathrm{rank}([\mathcal{I}_{xx} \;\; \mathcal{I}_{xu_0}]) = \tfrac{n(n+1)}{2} + nm$.
\EndFor
\State Estimate $\mathbb{E}_\mathcal{T}\{\mathcal{D}_{xx}\}$, $\mathbb{E}_\mathcal{T}\{\mathcal{I}_{xx}\}$, and $\mathbb{E}_\mathcal{T}\{\mathcal{I}_{xu_0}\}$ 
from offline data using \eqref{sample_estimate}--\eqref{sample_estimate3}.
\State \textit{Gain iteration:}
\For{$k = 0 : N_k$}
    \State Starting with a stabilizing $K_0$, compute $K$ using
    \begin{align}\label{eq:update ps}
        \underbrace{\begin{bmatrix}
        \mathbb{E}_\mathcal{T} \{\mathcal{D}_{xx}\}  & 
        -2\mathbb{E}_\mathcal{T}\{\mathcal{I}_{xx}(I \otimes K_k^T R)\} 
        -2\mathbb{E}_\mathcal{T}\{\mathcal{I}_{xu_0}(I \otimes R)\}
        \end{bmatrix}}_{\hat{\Theta}_k}
        \begin{bmatrix}
        \mathrm{vec}(P_k) \\ \mathrm{vec}(K_{k+1})
        \end{bmatrix} \nonumber \\
        = \underbrace{-\mathbb{E}_\mathcal{T}\{\mathcal{I}_{xx}\mathrm{vec}(Q_k)\}}_{\hat{\Phi}_k}.
    \end{align}
    \State Terminate when $\|P_k - P_{k-1}\| < \varsigma$, with $\varsigma > 0$ a small threshold.
\EndFor
\end{algorithmic}
\end{algorithm}

\normalsize
\begin{myth}
    With sufficiently large number of episodic trajectory under zero-mean process noise, using Algorithm 2 will generate a stabilizing LQR gain $K$ satisfying the following bound with respect to LQR gain for nominal dynamics without noise:
    \begin{align}
& \| K_{k+1} - K_{k+1}^{nom} \| \leq
\| \Theta_k^\dagger \|^2 \cdot   \| \Phi_k\| \cdot \gamma_{1\Delta}
+ \| \Theta_k^\dagger\| \| \Delta \mathcal{I}_{xx} \| \cdot \| \text{vec}(\bar{Q}_k) \| := \gamma_{2\Delta}, 
\end{align}
\normalsize
where, $\gamma_{1\Delta} = \| \Delta \mathcal{D}_{xx} \| + 2 \| \Delta \mathcal{I}_{xx} \| \cdot \| K_k^T R \| + 2 \| \Delta \mathcal{I}_{xu_0} \| \cdot \| R \|$. And the estimated solutions will converge
to a neighborhood of the optimal solution with the size of the neighborhood is determined by $\gamma_{2\Delta}$, and other notations follow from \eqref{difference_notation}.
\end{myth}
Proof: When the dynamic system is not perturbed with process noise, the model-free LQR gain computation is given by,
\begin{align}\label{eq:update_nominal}
 \underbrace{\begin{bmatrix}
\mathcal{D}_{xx}^{nom}  & -2\mathcal{I}_{xx}^{nom}(I \otimes K_k^{{nom}T}R)  -2\mathcal{I}^{nom}_{xu_0}(I \otimes R)  
\end{bmatrix}}_{\Theta_k^{nom}}
&\begin{bmatrix}
vec(P_k^{nom}) \\ vec(K_{k+1}^{nom}) 
\end{bmatrix}  =\underbrace{-\mathcal{I}^{nom}_{xx}vec(\bar{Q}_k)}_{\Phi_k^{nom}}.\\
\begin{bmatrix}
vec(P_k^{nom}) \\ vec(K_{k+1}^{nom}) 
\end{bmatrix} = \Theta_k^{nom\dagger}\Phi_k^{nom}
\end{align}

\normalsize
Then we consider the following perturbations:
\begin{align}
\label{difference_notation}
& \Delta \mathcal{D}_{xx} = \mathcal{D}_{xx} - \mathbb{E}_\mathcal{T} \{\mathcal{D}_{xx}\}, \quad \Delta \mathcal{I}_{xx} = \mathcal{I}_{xx} - \mathbb{E}_\mathcal{T} \{\mathcal{I}_{xx}\}, \nonumber \\
& \quad \Delta \mathcal{I}_{xu_0} = \mathcal{I}_{xu_0} - \mathbb{E}_\mathcal{T} \{\mathcal{I}_{xu_0}\} \\
&\Delta \Theta_k = \begin{bmatrix}
\Delta \mathcal{D}_{xx}  - 2 (\Delta \mathcal{I}_{xx})(I \otimes K_k^T R) - 2 (\Delta \mathcal{I}_{xu_0})(I \otimes R)
\end{bmatrix}
\end{align}
\[
\| \Delta \Theta_k \| \leq \| \Delta \mathcal{D}_{xx} \| + 2 \| \Delta \mathcal{I}_{xx} \| \cdot \| I \otimes K_k^T R \| + 2 \| \Delta \mathcal{I}_{xu_0} \| \cdot \| I \otimes R \|
\]
\normalsize 
Since \( \| I \otimes A \| = \| A \| \), we have,
\[
\| \Delta \Theta_k \| \leq \| \Delta \mathcal{D}_{xx} \| + 2 \| \Delta \mathcal{I}_{xx} \| \cdot \| K_k^T R \| + 2 \| \Delta \mathcal{I}_{xu_0} \| \cdot \| R \| := \gamma_{1\Delta}
\]
\[
\Delta \Phi_k = - (\mathcal{I}_{xx} - \mathbb{E}_\mathcal{T} \{\mathcal{I}_{xx}\}) \text{vec}(\bar{Q}_k)
\]
\normalsize

The norm is bounded as:
\[
\| \Delta \Phi_k \| = \| \mathcal{I}_{xx} - \mathbb{E}_\mathcal{T} \{\mathcal{I}_{xx}\} \| \cdot \| \text{vec}(\bar{Q}_k) \|
\]
\normalsize
Using perturbation theory for pseudoinverses:
\[
\| \Delta \Theta_k^\dagger \| \leq \| \Theta_k^\dagger \| \cdot \| \Delta \Theta_k \| \cdot \| \Theta_k^\dagger \|
\]
\normalsize

The perturbation in cost matrix and control gain can be bounded by,
\begin{align}
    \| \begin{bmatrix}
vec(P_k - P_k^{nom}) \\ vec(K_{k+1} - K_{k+1}^{nom}) 
\end{bmatrix} \| \leq \| \Delta \Theta_k^\dagger \|  \| \Phi_k\| + \| \Theta_k^\dagger\| \| \Delta \Phi_k \|,
\end{align}
\normalsize

neglecting $\|\Delta \Theta_k \Delta \Phi_k\|$. 
Therefore, the perturbation is bounded by,
\begin{align}
     \| \begin{bmatrix}
vec(P_k - P_k^{nom}) \\ vec(K_{k+1} - K_{k+1}^{nom}) 
\end{bmatrix} \| \leq \| \Theta_k^\dagger \|^2 \cdot \| \Delta \Theta_k \|  \| \Phi_k\| + \| \Theta_k^\dagger\| \| \Delta \Phi_k \|,\\
\leq \| \Theta_k^\dagger \|^2 \cdot   \| \Phi_k\| \cdot (\| \Delta \mathcal{D}_{xx} \| + 2 \| \Delta \mathcal{I}_{xx} \| \cdot \| K_k^T R \| + 2 \| \Delta \mathcal{I}_{xu_0} \| \cdot \| R \|) \nonumber \\ 
+ \| \Theta_k^\dagger\| \| \mathcal{I}_{xx} - \mathbb{E}_\mathcal{T} \{\mathcal{I}_{xx}\} \| \cdot \| \text{vec}(\bar{Q}_k) \| \\
\leq \| \Theta_k^\dagger \|^2 \cdot   \| \Phi_k\| \cdot \gamma_{1\Delta}
+ \| \Theta_k^\dagger\| \| \Delta \mathcal{I}_{xx} \| \cdot \| \text{vec}(\bar{Q}_k) \| := \gamma_{2\Delta}
\end{align}
\normalsize
The bounds $\gamma_{1\Delta}$ and $| \Delta \mathcal{I}_{xx} |$ capture perturbations arising from stochasticity in the dynamics and from the expectation operation. For the nominal system, the Kleinman update guarantees convergence of the matrices $P$ and $K$ to the optimal solution due to the monotonicity and contraction properties of the update operator. In particular, the sequence satisfies  $P_{opt}^{nom} \leq P_{k+1}^{nom} \leq P_{k}^{nom}$ where $P_{opt}^{nom} \succ 0$ is finite. When the offline data is persistently exciting, this ensures convergence to the optimal solution for the nominal system. With sufficiently large number of realizations, we would also have the error $\| \begin{bmatrix}
vec(P_k - P_k^{nom}) \\ vec(K_{k+1} - K_{k+1}^{nom}) 
\end{bmatrix} \|$ to shrink over iterative updates. Moreover, assuming the perturbation from nominal for the optimal solution is small, the error in the updates of the stochastic dynamic system $\| P_{k+1} - P_{opt}\|$ can be written as,
\begin{align}
    \| P_{k+1} - P_{opt}\| \leq \| (P_{k+1} - P_{k+1}^{nom})\| + \|(P_{k+1}^{nom} - P_{opt} ) \| \\
    \approx \| (P_{k+1} - P_{k+1}^{nom})\| + \|(P_{k+1}^{nom} - P_{opt}^{nom} ) \| ,\\
    \leq \gamma_{2\Delta} + \|(P_{k}^{nom} - P_{opt}^{nom} ) \|.
    % \approx \gamma_{2\Delta} + \|(P_{0}^{nom} - P_{opt}^{nom} ) \|
\end{align}
\normalsize
The stochastic error term $\gamma_{2\Delta}$ decays as the number of sample trajectories increases. Consequently, the estimated solutions converge monotonically to a neighborhood of the optimal solution under the Kleinman update, with the size of this neighborhood determined by the magnitude of $\gamma_{2\Delta}$.

\section{A Numerical Example}
\begin{figure}[H]
    \centering
\includegraphics[width=0.8\linewidth]{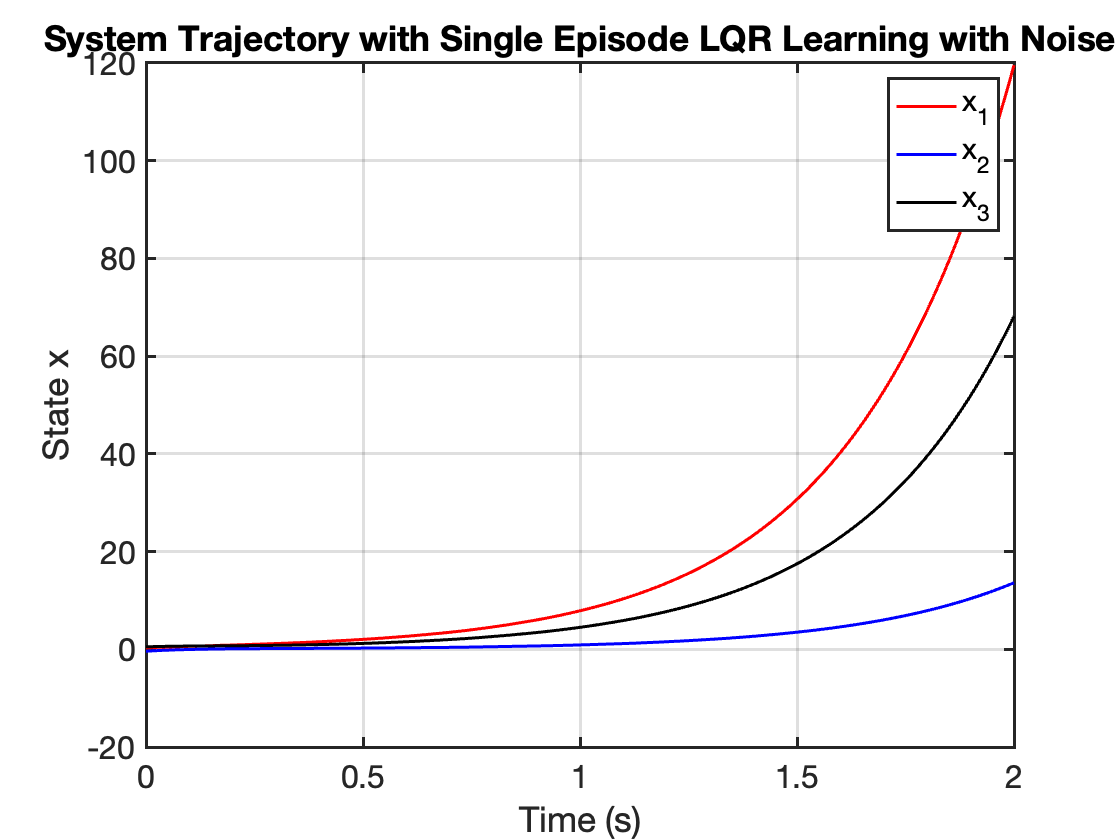}
    \caption{Single episode offline policy learning could not guarantee stability of the closed-loop uncertain dynamic system.}
    \label{fig:traj_unstable}
\end{figure}

We consider the following numerical example with the state, input, and disturbance matrices are considered to be as follows:
\[
A = \begin{bmatrix}
-2 & 1 & 0 \\
-4 & -5 & 0.4 \\
0 & -2 & -5
\end{bmatrix}, \quad
B = \begin{bmatrix}
1 \\
1 \\
1
\end{bmatrix}, \quad
E = \begin{bmatrix}
0.3 \\
0.3 \\
0.3
\end{bmatrix}
\]
Numerical example focuses on intricacies of learning the offline LQR for uncertain environment without any measurement of the exogenous inputs. With $Q=30I, R=1$, the model-based optimal without any disturbance input is $K = [4.0554,    1.0190,1.5051]$. First, we inject normally distributed signals (unknown to the designer), and gather state, and control input coupled trajectories $\mathcal{D}_{xx}, \mathcal{I}_{xx}, \mathcal{I}_{xu_0}$, where the exploration has been performed with sum of sinusoids $
u(t) = \sum_{i=1}^{N} \sin(\omega_i t)$ with frequencies are sampled uniformly from $\mathcal{U}(-5,5)$. When we use the gathered data to run the offline policy computation algorithm with a single episode, we lose the guarantees to obtain stable and convergent behavior, and often we get far from optimal and unstable policies. Fig. \ref{fig:traj_unstable} shows the behavior when the learned policy ($K = [-4.7022,6.2931 ,-1.1790]$) has been implemented in the dynamic system to test its performance. 

\begin{figure}[H]
    \centering
    \begin{subfigure}[b]{0.6\textwidth}
        \centering
    \includegraphics[width=\linewidth]{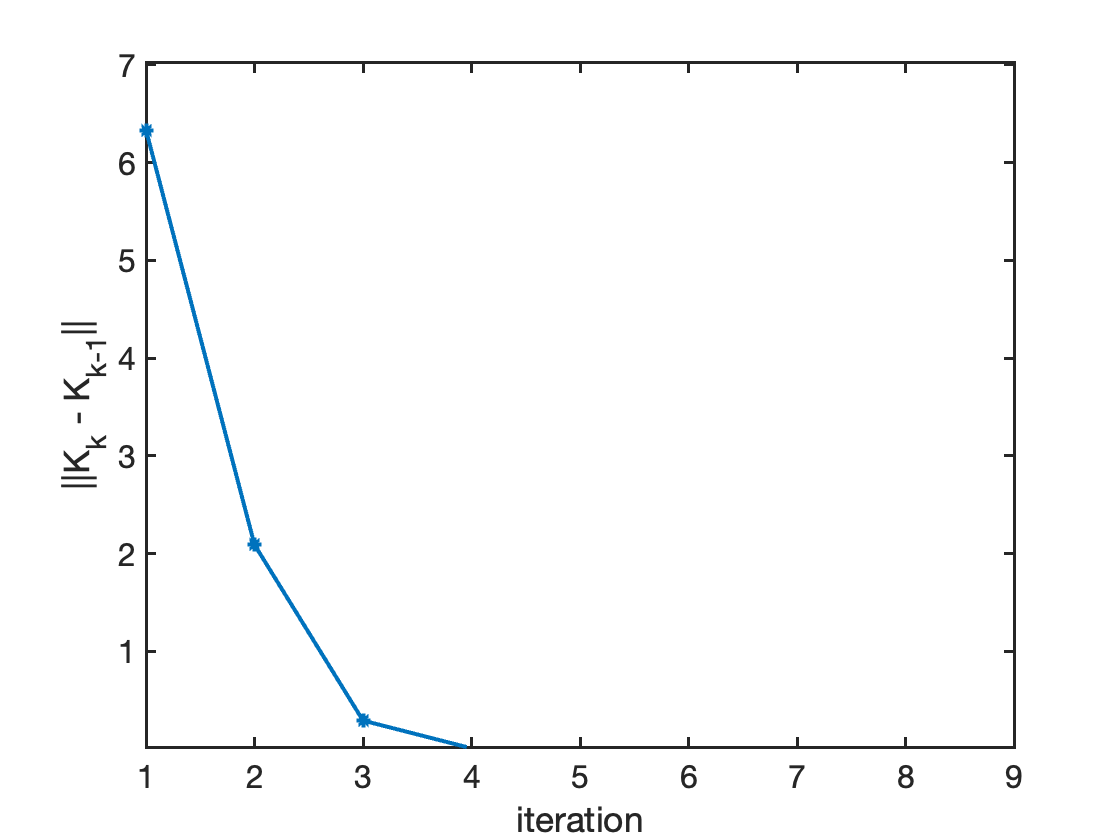}
        \caption{Convergence of offline LQR gain $K$ with the episodic Algorithm 2 under unmeasured uncertainty in state dynamics.}
        \label{fig:fig1}
    \end{subfigure}
    \vfill
    \begin{subfigure}[t]{0.6\textwidth}
        \centering
\includegraphics[width=\linewidth]{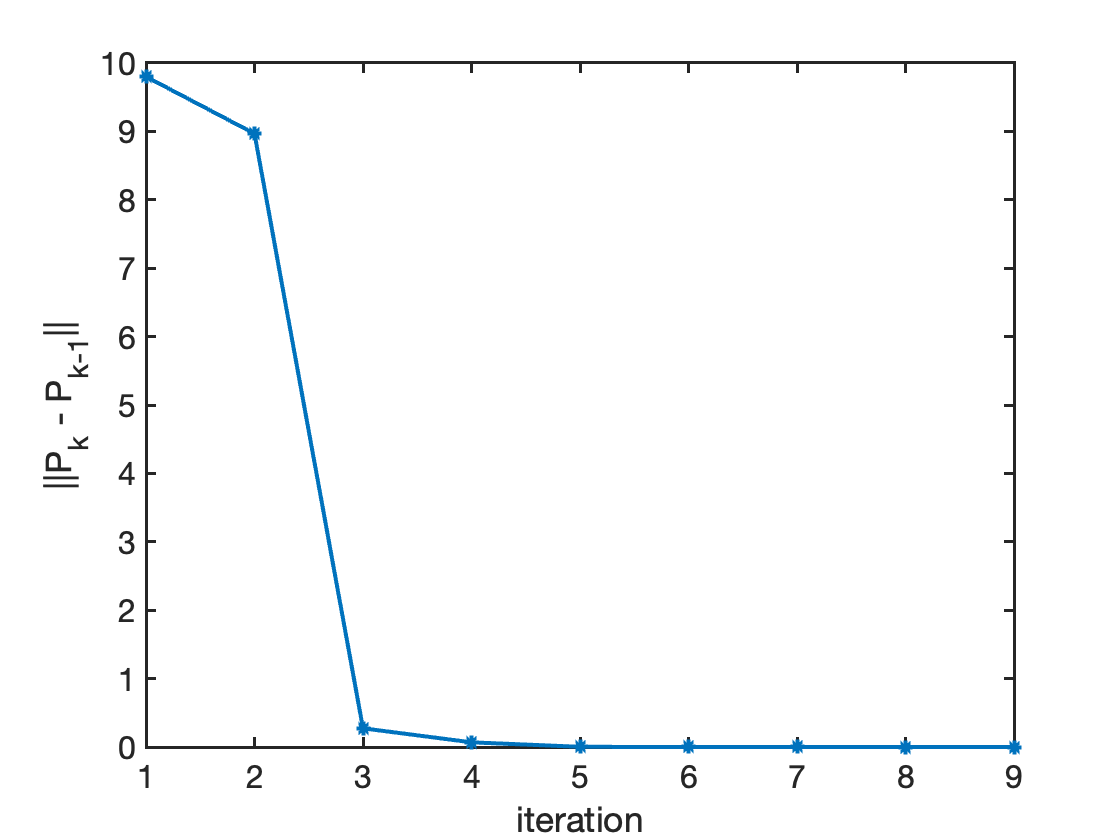}
        \caption{Convergence of offline Riccati solution $P$ with the episodic Algorithm 2 under unmeasured uncertainty in state dynamics.}
        \label{fig:fig2}
    \end{subfigure}
    \caption{Convergence performance of the offline episodic policy computation algorithm for the system under unmeasured uncertainty.}
    \label{fig:convergence}
\end{figure}

For one single episode, data has been gathered for $100$ samples to satisfy the persistency of excitation condition that requires at-least $n(n+1)/2 + nm$ samples. We then consider multiple realizations of the oracle based exploration as prescribed in Algorithm 2. We consider $50$ realizations or sample paths, and subsequently utilize constructed expected data matrices $\mathbb{E}_\mathcal{T} \{\mathcal{D}_{xx} \}, \mathbb{E}_\mathcal{T} \{\mathcal{I}_{xx} \}, \mathbb{E}_\mathcal{T} \{\mathcal{I}_{xu_0} \}$ to compute the LQR update iterations for $P_k$, and $K_k$, $k=1,2,\dots$ with gain computed as $K = [4.0561,1.0218,1.5112$]. The convergences of them are shown in Fig. \ref{fig:convergence}, that shows quite fast convergence within $5$ iteration steps.  Furthermore, when we implement this offline learned gain to the closed-loop analysis in presence of stochasticity, then system states show sufficiently fast convergence to origin with trajectories remain close to the steady state neighborhood with the process noise.  

\begin{figure}
\centering
\includegraphics[width=0.8\linewidth]{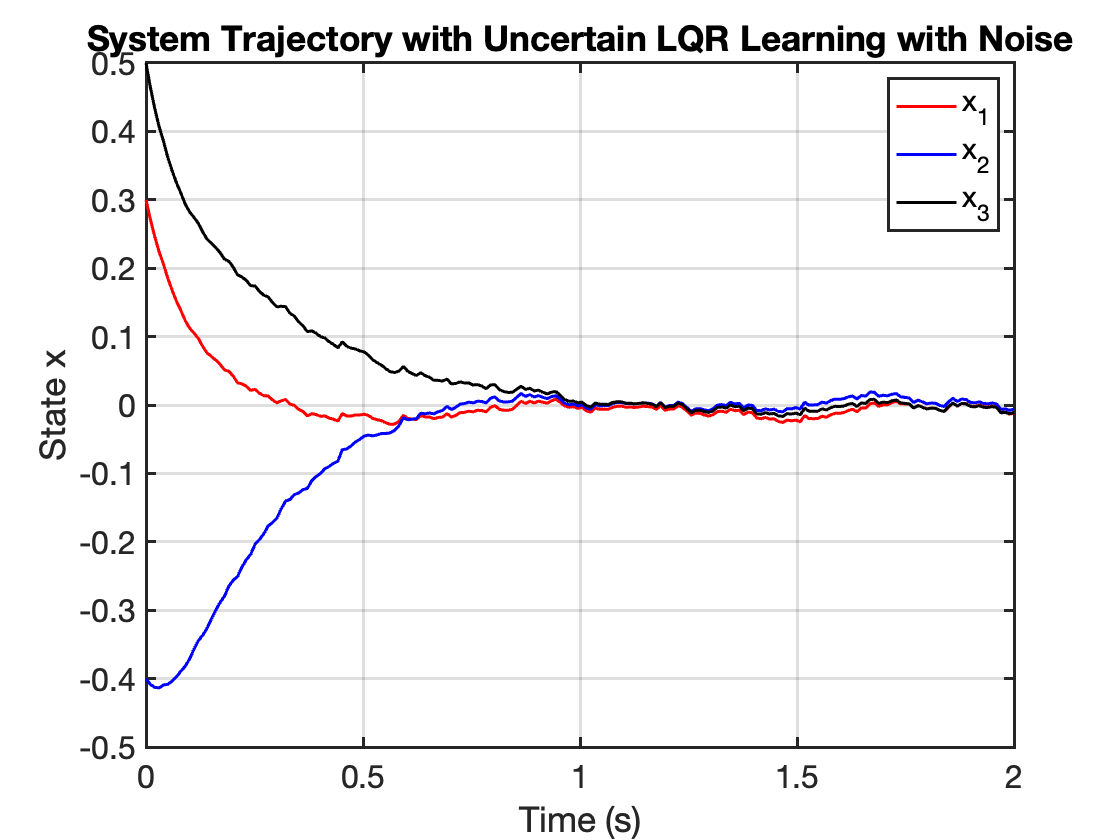}
\label{fig:traj_stable}
    \caption{Episodic offline policy learning with Algorithm 2 guarantees stability of the closed-loop uncertain dynamic system.}
    \label{fig:side_by_side}
\end{figure}

% \begin{figure}
%     \centering
%     \includegraphics[width=0.5\linewidth]{Figures/K_iter.png}
%     \caption{Caption}
%     \label{fig:enter-label}
% \end{figure}

\vspace{-0.6 cm}
\section{Conclusions}
This paper presents a system-theoretic development and analysis of offline LQR learning in presence of exogenous disturbances. We discussed two different scenarios. In the first scenario the exogenous inputs are measurable in a controlled environment leading to an error-compensating offline learning with one-shot trajectory data. Although as for most of the systems, disturbances are difficult to estimate, moreover with unknown dynamics, we then formulated an episodic approximated ADP-based gain computation framework. Stability and convergence analysis have been performed that shows sufficient guarantees that the gains will converge near to the optima with the neighborhood radius dictated by the stochasticity encountered by the system.  

\section*{Acknowledgment}

This research was supported by the U.S. Department of
Energy, through the Office of Advanced Scientific Computing Research’s “Data-Driven Decision Control for Complex
Systems (DnC2S)” project. Pacific Northwest National Laboratory is operated by Battelle Memorial Institute for the
U.S. Department of Energy under Contract No. DE-AC05-
76RL01830. Authors would like to thank Dr. Subhrajit Sinha at PNNL for helpful discussions. 

\vspace{-0.4 cm}

\bibliographystyle{IEEEtran}
\bibliography{Refs}
\end{document}